\documentclass[10pt,final,doublecolumn]{IEEEtran}
\usepackage{amsmath}
\usepackage{amssymb}
\usepackage{amsfonts}
\usepackage{latexsym}
\usepackage{graphicx}
\usepackage{bbding}
\usepackage{enumerate}
\usepackage{subeqnarray}
\usepackage{multicol}
\usepackage{color}
\usepackage{slashbox}
\usepackage{setspace}
\usepackage{mathrsfs}
\usepackage{algorithm,algorithmic}
\usepackage{subfigure}
\usepackage{algorithmic}
\usepackage{algorithm,float}
\usepackage{makecell}
\usepackage{bm}
\IEEEoverridecommandlockouts

\makeatletter
\newenvironment{breakablealgorithm}
{
		\begin{center}
			\refstepcounter{algorithm}
			\hrule height.8pt depth0pt \kern2pt
			\renewcommand{\caption}[2][\relax]{
				{\raggedright\textbf{\ALG@name~\thealgorithm} ##2\par}%
				\ifx\relax##1\relax 
				\addcontentsline{loa}{algorithm}{\protect\numberline{\thealgorithm}##2}%
				\else 
				\addcontentsline{loa}{algorithm}{\protect\numberline{\thealgorithm}##1}%
				\fi
				\kern2pt\hrule\kern2pt
			}
		}{
		\kern2pt\hrule\relax
	\end{center}
}
\makeatother

\begin{document}

\title{Exploiting Tensor-based Bayesian Learning for Massive Grant-Free Random Access in LEO Satellite Internet of Things}
\author{\IEEEauthorblockN{
Ming Ying, Xiaoming Chen, and Xiaodan Shao
\thanks{Ming Ying (Email: {\tt 3180103509@zju.edu.cn}) and Xiaoming Chen (Email: {\tt chen\_xiaoming@zju.edu.cn}) are with the College of Information Science and Electronic Engineering, Zhejiang University, Hangzhou 310027, China. Xiaodan Shao (Email: {\tt shaoxiaodan@zju.edu.cn}) is with the Institute for Digital Communications (IDC), Friedrich-Alexander University of Erlangen-Nuremberg, 91058 Erlangen, Germany.}}}
\maketitle

\begin{abstract}
With the rapid development of Internet of Things (IoT), low earth orbit (LEO) satellite IoT is expected to provide low power, massive connectivity and wide coverage IoT applications. In this context, this paper provides a massive grant-free random access (GF-RA) scheme for LEO satellite IoT. This scheme does not need to change the transceiver, but transforms the received signal to a tensor decomposition form. By exploiting the characteristics of the tensor structure, a Bayesian learning algorithm for joint active device detection and channel estimation during massive GF-RA is designed. Theoretical analysis shows that the proposed algorithm has fast convergence and low complexity. Finally, extensive simulation results confirm its better performance in terms of error probability for active device detection and normalized mean square error for channel estimation over baseline algorithms in LEO satellite IoT. Especially, it is found that the proposed algorithm requires short preamble sequences and support massive connectivity with a low power, which is appealing to LEO satellite IoT.
\end{abstract}

\begin{IEEEkeywords}
Internet of Things, grant-free random access, low earth orbit satellite, Bayesian learning.
\end{IEEEkeywords}

\section{Introduction}
Nowadays, Internet of Things (IoT) has been widely applied in various fields, e.g., industry, agriculture, traffic and medicine \cite{IoT1}-\cite{IoT3}. As a typical machine-type communication, IoT have two important characteristics compared to traditional human-type communication. The first one is massive connectivity. It is predicted that in 2025, the number of IoT devices will reach 21.5 billion. In this context, massive machine-type communication (mMTC) has been defined as one of main use cases of 5G wireless networks, namely 5G cellular IoT \cite{CIOT1}. The second one is wide coverage. IoT has been applied not only in urban areas, but also in remote areas, e.g., forest, ocean, mountain, and desert. However, these remote areas do not have effective terrestrial wireless coverage. To this end, low earth orbit (LEO) satellite IoT is proposed and receives considerable interests due to short propagation latency and small path loss \cite{Reomte}. In recent years, SpaceX and OneWeb companies launch a large number of LEO satellites to provide global coverage \cite{Spacex} \cite{Oneweb}.

Considering the bursty nature of IoT applications, random access protocol is adopted in IoT to save the energy \cite{RA}. For instance, 5G cellular IoT employs the famous ALOHA protocol \cite{CIOT2}. Meanwhile, various improved ALOHA protocols are also widely utilized in traditional satellite communications \cite{ALOHA1} \cite{ALOHA2}. {  ALOHA is a commonly used grant-based random access (GB-RA) protocol, which requires four transmissions between access point and active devices \cite{A1}. Firstly, each active device randomly selects a preamble sequence from a set of orthogonal sequences and sends the sequence to the access point. Next, the access point responds to each active device, authorizing them to send connection requests. After that, the active device sends a connection request to the access point for resource allocation to transmit data. Finally, if the preamble sequence that the active device sends is unique, the access point will authorize the corresponding request and send a contention-resolution message to inform the active device of the available resources}. 
For LEO satellite IoT, due to long transmission distance (from 400 to 2000 kilometers), four transmissions lead to a high access latency. Especially in the scenario of massive connectivity, {ALOHA} may have a high access failure probability, which further increases the access latency. {Moreover, the ASL spacemobilie launched a LEO satellite named bluewalker 3 recently. Such a LEO satellite can support the direct access of a massive number of mobile devices, which is an important trend of LEO satellite IoT. In the context of massive direct access, GF-RA is a promising protocol.} To this end, grant-free random access protocol is introduced to LEO satellite IoT \cite{SGFRA1} \cite{SGFRA2}. Specifically, after sending their assigned preamble sequences, active devices transmit their data signals directly without the grant of LEO satellites. Thus, the access latency can be decreased significantly. Therefore, grant-free random access is appealing to LEO satellite IoT.

The key of grant-free random access is to detect active devices from the received preamble sequences \cite{GFRA1}-\cite{GFRA3}. Since the preamble sequences are not orthogonal in the scenario of massive connectivity, active device detection is not trivial. Considering only a small portion of devices are active in a time slot due to the bursty nature of IoT applications, active device detection is usually formulated as a compressed sensing problem. For such a problem, approximate message passing (AMP) is an effective approach \cite{AMP1} \cite{AMP2}. In \cite{AMP3}, the authors analyzed the activity detection performance of AMP in cellular IoT. It is proved that as the number of base station antennas tends to infinity, the activity error probability of AMP asymptotically approaches zero. However, if the number of base station antennas is limited, AMP requires long preamble sequences in order to guarantee the accuracy of active device detection. With the goal of decreasing the required length of preamble sequences, covariance-based approaches are applied to active device detection \cite{Covariance1} \cite{Covariance2}. Specifically, the covariance information of the received signal is utilized to detect the active devices. In \cite{Covariance3}, over Rayleigh fading channels, the authors proposed an active device detection algorithm by maximizing the likelihood function of the received signal. Furthermore, a joint activity detection and channel estimation algorithm based on the covariance of the received signal was designed in \cite{Covariance4}. Moreover, optimization-based approaches also can be used to detect the active devices. In \cite{Opt}, the authors first projected the received signal to a low-dimension space, and then employed a Riemann optimization method to judge the active devices. {In \cite{shaotensor}, the authors proposed a new reconfigurable intelligent surface-aided massive access architecture and formulated joint active device separation and channel estimation as a coupled high-order tensor problem, which was addressed by using a Bayesian learning method \cite{shaosensing}.}

A common assumption to the above active device detection algorithms for grant-free random access is that the channels experience Rayleigh fading. Yet, for LEO  satellite IoT, due to the existence of light-of-sight (LOS) transmission, the direct application of the above detection algorithms may lead to severe performance degradation. To the best of the authors' knowledge, grant-free random access in LEO satellite IoT is still an open issue. Recently, tensor-based approaches are applied in unsourced random access \cite{TRA} \cite{TRA2}. It is shown that such approaches are able to recover data codewords from the mixed received signals exactly. {However, these works as \cite{TRA2} are all proposed for Rayleigh channel, which may not be applicable in the scenarios of Rician channel.} In this context, this paper intends to design a simple but effective tensor-based grant-free random access scheme for LEO satellite IoT in presence of LOS transmission. The contributions of this paper are as follows.

\begin{enumerate}

\item We propose a novel framework of massive grant-free random access for LEO satellite IoT. Such a framework does not need to change the transceiver of LEO satellite IoT, but only transforms the received signal to a tensor decomposition form.

\item We design a {low-complexity} joint activity detection and channel estimation algorithm based on the proposed massive grant-free random access framework by exploiting the tensor structure of the received signal.

\item We analyze the convergence behavior and computational complexity of the proposed algorithm, and verify the effectiveness of the proposed algorithm in LEO satellite IoT via extensive simulations.

\end{enumerate}

The rest of this paper is organized as follows. In Section II, we introduce the considered LEO satellite IoT network with the focus on the adopted massive grant-free random access protocol. Then, we propose a tensor-based Bayesian learning algorithm for joint activity detection and channel estimation, and analyze the convergence and complexity of the proposed algorithm in Section III. After that, we present extensive simulation results in Section IV to evaluate the performance of the proposed algorithm. Finally, Section V concludes the paper.

\emph{Notations:} We use bold upper (lower) letters to denote matrices (column vectors), non-bold letters to denote scalars, $\mathbb{C}^{X \times Y}$ to denote the space of complex matrices of size $X\times Y$, $(\cdot)^H$ and $(\cdot)^T$ to denote conjugate transpose and transpose,  $*$ to denote conjugation, $\mathrm{Tr}(\cdot)$ to denote the trace of a matrix, $\mathrm{diag}(\mathbf{a})$ to denote a diagonal matrix with the diagonal entries specified by vector $\mathbf{a}$, $\mathbf{I}_K$ to denote a $K \times K$ identity matrix, $\mathbf{1}_K $ to denote an all-one vector with length $K$, $\mathrm{vec}(\cdot)$ to denote column vectorization, $\Vert \cdot\Vert_F$ to denote Frobenius-norm of a matrix, $\left[\!\left[\cdot\right]\!\right]$ to denote the Kruskal operator, $\otimes$ to denote the Kronecker product, $\circ$ to denote the vector outer product, $\odot$ to denote the Hadamard product, $\diamond$ to denote the Khatri-Rao product, $\mathcal{CN}(x|\mu,\sigma^2)$ to denote complex Gaussian distribution with mean $\mu$ and variance $\sigma^2$,  $\mathcal{U}$ to denote uniform distribution, $p(\cdot|\cdot)$ to denote conditional probability distribution, $\mathbb{E}$ to denote the expectation of a variable, $\mathrm{Gamma}(\cdot)$ to denote Gamma function, $\mathrm{Gamma}(x|\alpha,\beta)$ to denote the variable $x$ obeying Gamma distribution with parameters $\alpha$ and $\beta$, $\mathrm{Hy}(\cdot)$ to denote Confluent Hypergeometric Function. For a matrix $\mathbf{A}$, we use $\mathbf{A}(a,b)$ to denote its $(a,b)$-th element, $\mathbf{A}(:,k)$ and $\mathbf{A}(k,:)$ to denote its $k$-th column and $k$-th row, respectively. {Moreover, we give out the following definition, which is used in the rest of this paper.}

{\emph{Definition 1:} The Confluent Hypergeometric Function $\mathrm{Hy}(a,b,x)$ for all real or complex $a$, $c$, $x$, is given by the power series \cite{E1}
	\begin{equation}\label{Hy}
		\mathrm{Hy}(a,b,x)=\sum_{v=0}^{\infty}\frac{(a)_v}{(c)_v}\frac{x^v}{v!}
	\end{equation}
	where $(a)_v=a(a+1)\cdots(a+v-1),(a)_0=1,(1)_v=v!,(v=0,1,2,\cdots) $ is the Pochhammer's symbol.}

\section{System Model}

We consider a LEO satellite IoT network as shown in Fig. \ref{channel_model}, where a LEO satellite equipped  with $M$ antennas\footnote{Note that multiple antennas are commonly used in current LEO satellites \cite{F2}, \cite{F1}.} serves $K$ single-antenna IoT devices distributed over a large area. Due to the bursty characteristics of IoT applications, only a small portion of devices have data to send in a time slot. In order to decrease the access latency in the scenario of long distance between LEO satellite and IoT devices, a grant-free random access (GF-RA) protocol is employed in the LEO satellite IoT. Specifically, each IoT device is assigned a unique preamble sequence. At the beginning of each time slot, active devices transmit their preamble sequences to inform the satellite that they have data to send. Based on the received signal, the LEO satellite detects the active devices and estimates their corresponding channel state information (CSI), which is used for data signal decoding in the rest of the time slot. In what follows, we introduce the considered LEO satellite channel model and the adopted GF-RA protocol, respectively.


	
	
	
\subsection{Channel Model}

\begin{figure}[!h]
	\centering
	\includegraphics [width=0.5\textwidth] {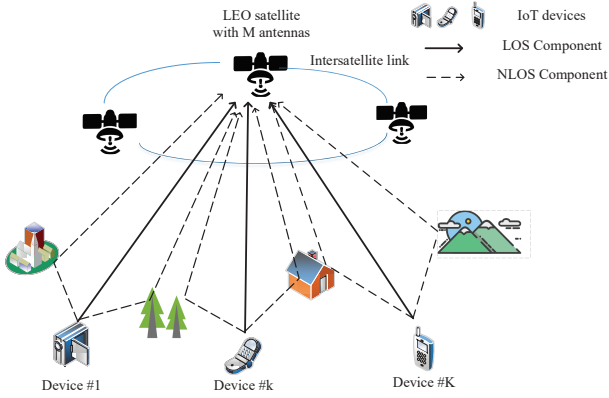}
	\caption {System model of LEO satellite IoT network. }
	\label{channel_model}
\end{figure}
According to the signal propagation characteristics of LEO satellite communications {\cite{SGFRA1}, \cite{SGFRA2}, \cite{F2}-\cite{F6}}, the LEO satellite channel usually includes two components, namely line of sight (LOS) and non-line of sight (NLOS), as shown in Fig. \ref{channel_model}. Hence, the channel {in the commonly used Ka band} between device $k$ and the LEO satellite can be expressed as
\begin{equation}\label{channel1}
	\mathbf{h}_k=\omega_k\bigg(\sqrt{\frac{\lambda_k g_k}{\lambda_k+1}}\mathbf{h}^{LOS}_k+\sqrt{\frac{g_k}{\lambda_k+1}}\mathbf{h}^{NLOS}_k\bigg),
\end{equation}
{where $g_k$ is the large-scale fading factor of the channel between IoT device $k$ and the LEO satellite, given by \cite{F4}, \cite{F5}}
\begin{equation}\label{g_k}
	 g_k = (\frac{c}{4\pi fd_0})^2\cdot \frac{G_k}{\kappa BT}\cdot \frac{1}{r_k},
\end{equation}
{where $(\frac{c}{4\pi fd_0})^2$ is the free space loss (FPL) with $c$ being the light speed, $f$ being the carrier frequency, $d_0$ being the propagation distance, $G_k$ being the transmit antenna gain of the device $k$, $\kappa$ being the Boltzman's constant, $B$ being the carrier bandwidth, $T$ being the temperature of the received noise and $r_k$ is the rain attenuation cofficient of device $k$ whose power gain in dB $r_n^{dB}=20\log_{10} r_n$ , follows log-normal random distribution $\ln(r_n^{dB})\sim\mathcal{N}(\mu_r, \sigma_r^2)$ \cite{F6}. Moreover, $\omega_k$ is the satellite receive annenna gain, which is given by \cite{F3}
	\begin{equation}\label{omega_k}
		\omega_k=\frac{J_1(\phi_k)}{2\phi_k}+36 \frac{J_3(\phi_k)}{\phi_k^3},
	\end{equation}
where $\phi_k = \frac{\pi d_sf}{c}\sin(\theta_k)$ with $d_s$ being the dianeter of circular antenna array on the satellite and $\theta_k$ being the off-axis of the satellite boresight to device $k$. For channel small-scale fading, $\lambda_k $ is the Rician factor}, $\mathbf{h}_k^{LOS}\in {\mathbb{C}}^{1\times M}$ is the LOS component of LEO satellite channel, which can be seen as a constant in a relatively long time since IoT devices are usually deployed at fixed position and their transmit elevation angles to the LEO satellite keep invariant. On the other hand, ${\mathbf{h}_k^{NLOS}\in {\mathbb{C}}^{1\times M}}$  is the NLOS component of LEO satellite channel, which follows the i.i.d. complex Gaussian distribution, i.e., $\mathbf{h}_k^{NLOS}\sim \mathcal{CN}(\mathbf{0},v_k^{NLOS}\mathbf{I}_M)$ with $v_k^{NLOS}$ being the variance. Note that the channel parameters $\mathbf{h}_k^{LOS}\in {\mathbb{C}}^{1\times M}$ and $v_k^{NLOS}$ are related to carrier frequency, the link conditions, and so on \cite{channel2}. Therefore, the channel $\mathbf{h}_k\in {\mathbb{C}}^{1\times M}$ can be regarded as the combination of the LOS component $\mathbf{h}_k^{LOS}$ and the NLOS component $\mathbf{h}_k^{NLOS}$, with the distribution $\mathbf{h}_k\sim \mathcal{CN}(\omega_k\sqrt{\frac{\lambda_kg_k}{\lambda_k+1}}\mathbf{h}_k^{LOS}, \frac{\omega_k^2g_k}{\lambda_k+1}v_k^{NLOS}\mathbf{I}_M)$.

\subsection{GF-RA Protocol}
Considering the above channel characteristics of LEO satellite IoT, we propose a tensor-based GF-RA protocol.
 {First, a unique preamble sequence of length $L$ is designed for each IoT device. Without loss of generality, it is assumed that the length $L$ of preamble sequences can be factorized as $L=\prod_{i=1}^{d}l_i$ for some $d\geq2$, $l_i\geq2,i=1,2,\cdots ,d$. Subsequently, a rank-1 tensor $\mathcal{A}_k$ of dimensions $l_1, l_2, \cdots,l_d$ is generated as
	\begin{equation}\label{rank1}
		\mathcal{A}_k=\mathbf{a}_{1,k}\circ\mathbf{a}_{2,k}\circ\cdots\circ\mathbf{a}_{d,k}, \quad\forall k
	\end{equation}
	where $\mathbf{a}_{i,k}\in\mathbb{C}^{l_i\times 1}, i\in\{1,2\cdots d\}$ is a series of vectors Gaussian distributed with unit norm. Then, the preamble sequence $\mathbf{a}_k$  for device $k$ is constructed as
	\begin{equation}\label{piotse}
		\mathbf{a}_k = \text{vec}(\mathcal{A}_k)\in\mathbb{C}^{\prod_{i=1}^{d}l_i}=\mathbb{C}^{L\times 1}.
\end{equation}}
According to the property of outer product \cite{TRA}, we have
\begin{equation}\label{ak}
	\mathbf{a}_k=\mathbf{a}_{1,k}\otimes\mathbf{a}_{2,k}\otimes\cdots\otimes\mathbf{a}_{d,k}.
\end{equation}
Once device $k$ is activated, $\mathbf{a}_k$ is sent to the LEO satellite at the beginning of the time slot for joint activity detection and channel estimation (JADCE). Thus, the received signal $\mathbf{Y}\in\mathbb{C}^{L\times M}$ at the LEO satellite can be expressed as\footnote{The large carrier frequency
	offset (CFO) caused by high mobility of the LEO satellite is compensated before according to the deterministic LEO's trajectory [37], [38].}
\begin{equation}\label{matrixy}
	\mathbf{Y} = \sum_{k=1}^{K}\mathbf{a}_{k}{\alpha_{k}\sqrt{\xi_k}\mathbf{h}_k^H}+\mathbf{N}=\sum_{k=1}^{K}\mathbf{a}_{k}\mathbf{x}_k^H+\mathbf{N},
\end{equation}
where  $\mathbf{h}_k$ is the channel between device $k$ and the LEO satellite described in (\ref{channel1}), ${\xi_k}$ is the transmit power of preamble sequence, $\mathbf{N}\in\mathbb{C}^{L \times M}$ is the additive white Gaussian noise with variance $\sigma_n^2$, and $\alpha _k$ is the activity indicator with $\alpha _k = 1$ if the $k$-th device is active and $\alpha _k = 0$ otherwise. Considering the activity possibility $p_k$, we have
\begin{equation}\label{pk}
	\begin{cases}
		\mathrm{Pr}(\alpha _k = 1) = p_k   \\
		\mathrm{Pr}(\alpha _k  = 0) = 1-p_k
	\end{cases} .
\end{equation}
For simplicity, we define $\mathbf{x}_k = \alpha_k\sqrt{\xi_k}\mathbf{h}_k^H\in\mathbb{C}^{M\times 1}$ in (\ref{matrixy}) as the device state vector of the device $k$.

With the received signal $\mathbf{Y}$, the LEO satellite transforms it to a vectorized form with the Kronecker product as
\begin{equation}\label{vectorized}
	\mathbf{y}=\sum_{k=1}^{K}\mathbf{a}_k\otimes{\mathbf{x}_k}+\mathbf{n},
\end{equation}
where $\mathbf{y}\in \mathbb{C}^{LM}$ and $\mathbf{n}\in \mathbb{C}^{LM}$ denote the vectorized versions of $\mathbf{Y}$ and $\mathbf{N}$, respectively. Substituting (\ref{ak}) to (\ref{vectorized}), the vectorized received signal $\mathbf{y}$ can be rewritten in terms of vectors $\mathbf{a}_{i,k}\in\mathbb{C}^{l_i}$,$1\leq i\leq d $ as follows
\begin{equation}\label{vectorizedy}
	\mathbf{y}=\sum_{k=1}^{K}\mathbf{a}_{1,k}\otimes\mathbf{a}_{2,k}\otimes\cdots\otimes\mathbf{a}_{d,k}\otimes{\mathbf{x}_k}+\mathbf{n}.
\end{equation}
Further, the LEO satellite rearranges the vectorized received signal $\mathbf{y}$ into the tensor decomposition form $\mathcal{Y}\in\mathbb{C}^{l_1\times l_2\times \cdots\times l_d\times M}$ as
\begin{equation}\label{tensory}
	\mathcal{Y} = \sum_{k=1}^{K}\mathbf{a}_{1,k}\circ\mathbf{a}_{2,k}\circ\cdots\circ\mathbf{a}_{d,k}\circ \mathbf{x}_k+\mathcal{N},
\end{equation}
where $\mathcal{Y}$ is the received signal in the tensor space, and $\mathcal{N}\in\mathbb{C}^{l_1\times l_2\times \cdots \times l_d\times M}$ is the additive white Gaussian noise in the same tensor space.

It is clear that the key of JADCE is to recover the device state vector $\mathbf{x}_k$ from the received signal $\mathcal{Y}$. Then, the activity indicator and CSI can be acquired based on the recovered $\mathbf{x}_k$. In the next section, according to the characteristics and requirements of LEO satellite IoT, we design a simple but effective JADCE algorithm by exploiting the tensor structure of the received signal.

\section{Tensor-based Bayesian Learning for JADCE}
In this section, we aim to design a JADCE algorithm for LEO satellite IoT based on the mixed received signal. Considering the tensor decomposition form of the received signal $\mathcal{Y}$ in (\ref{tensory}), we can formulate JADCE as the following optimization problem
\begin{equation}\label{problem1}
	\begin{aligned}
		&\mathop{\arg\min}_{{\mathbf{x}_k\in\mathbb{C}^M}}\quad\Bigg\Vert \mathcal{Y}-\sum_{k=1}^{K}\mathbf{a}_{1,k}\circ\mathbf{a}_{2,k}\circ \cdots \circ\mathbf{a}_{d,k}\circ{\mathbf{x}_k}\Bigg\Vert_{F}^2 	\\
		        	\mathrm{s.t.} &	 \sum_{k=1}^{K}\big\Vert\mathbf{x}_k^H\mathbf{x}_k\big\Vert_0\leq \delta_0,
	\end{aligned}.
\end{equation}
{ where $\delta_0$ is a predefined parameter for imposing the channel sparsity.} In order to simplify the expression of  problem (\ref{problem1}), we adopt the Kruscal operator $\left[\!\left[\cdot\right]\!\right]$ and factor matrices $\mathbf{A}_1,\mathbf{A}_2,\cdots\mathbf{A}_d,\mathbf{X}$. Then, we have

\begin{equation}\label{problem2}
	\begin{aligned}
		&\mathop{\arg\min}_{{\mathbf{X}\in\mathbb{C}^{M\times K}}}\quad\big\Vert \mathcal{Y}-\left[\!\left[\mathbf{A}_1,\mathbf{A}_2,\cdots\mathbf{A}_d,\mathbf{X}\right]\!\right]\big\Vert_{F}^2 	\\
				\mathrm{s.t.} &\sum_{k=1}^{K}\big\Vert\mathbf{X}(:,k)^H\mathbf{X}(:,k)\big\Vert_0\leq \delta_0 \\
	\end{aligned},
\end{equation}
where $\mathbf{A}_i=[\mathbf{a}_{i,1},\mathbf{a}_{i,2},$ $\cdots, \mathbf{a}_{i,K}]\in\mathbb{C}^{l_i\times K}$, $i=1,2,\cdots,d$ with the $k$-th column being $\mathbf{a}_{i,k}$, and $\mathbf{X}=[\mathbf{x}_{1},\mathbf{x}_{2},$ $\cdots, \mathbf{x}_{K}]\in\mathbb{C}^{M\times K}$ with the $k$-th column being $\mathbf{x}_{k}$. To handle the JADCE problem in (\ref{problem2}), we design an intelligent algorithm that can automatically learn the device state matrix $\mathbf{X}$ from the received signal in the tensor space at the LEO satellite by using a Bayesian learning approach.

\subsection{Probabilistic Modeling}

\begin{figure}[!h]
	\centering
	\includegraphics [width=0.4\textwidth] {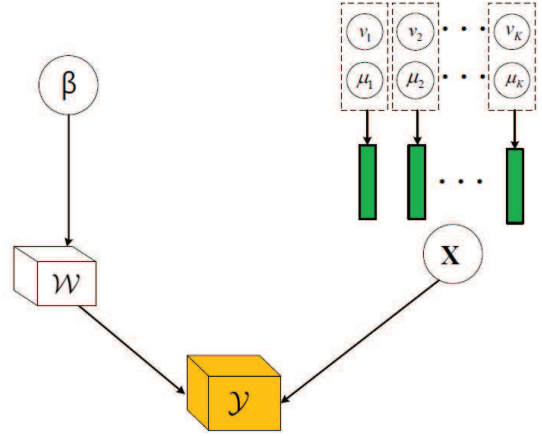}
	\caption {{The orange cube denotes the observable variables, especially the received signal $\mathcal{Y}$ in tensor form, white cube denotes the AWGN $\mathcal{W}$ in tensor form, and white circle denotes $\beta$.} The green rectangles denote the rank-1 tensor of factor matrix $\mathbf{X}$, and arrows describe conditional dependencies between variables. }
	\label{promodel}
\end{figure}

In order to apply the Bayesian learning approach, we shall build a probabilistic model for problem (\ref{problem2}). As shown in Fig. \ref{promodel}, we construct the probabilistic model by using some probability density functions (pdfs) to interpret each unknown term in the problem (\ref{problem2}). Firstly, let us investigate the device state matrix $\mathbf{X}$. According to the LEO satellite channel characteristics in (1) and the activity probability in (3), { especially the LOS component $\mathbf{h}_k^{LOS} $ and the NLOS component $\mathbf{h}_k^{NLOS}$}, $\mathbf{X}$ can be modelled as the following circularly-symmetric complex Gaussian prior distribution for each column of $\mathbf{X}$
\begin{equation}\label{pG}
	p\Big(\mathbf{X}|\{\mu_{k}\}_{k=1}^K , \{v_{k}\}_{k=1}^K\Big) = \prod_{k=1}^{K} \mathcal{CN}(\mathbf{X}(:,k)|\mu_{k}^{-1}\mathbf{1}_M,v_k^{-1}\mathbf{I}_M),
\end{equation}
where $\mu_k^{-1}$ and $v_k^{-1}$ are mean and variance for the LEO satellite channel of device $k$, respectively. Considering low activity probabilities of IoT devices, the device state matrix $\mathbf{X}$ is usually of column sparsity. In order to promote sparsity, we adopt gamma distribution to characterize the parameters $\mu_k$ and $v_k$ in (\ref{pG}) as below \cite{convp}
\begin{equation}\label{pmuk}
	p\big(\{\mu_{k}\}_{k=1}^K|\mathbf{i}_{\mu}\big) = \prod_{k=1}^{K} \mathrm{Gamma}(\mu_{k}|\varepsilon,\varepsilon)= \prod_{k=1}^{K} \mu_{k}^{\varepsilon-1}\mathrm{exp}(-\varepsilon\mu_{k}),
\end{equation}
\begin{equation}\label{pvk}
	p\big(\{v_{k}\}_{k=1}^K|\mathbf{i}_{v}\big) = \prod_{k=1}^{K} \mathrm{Gamma}(v_{k}|\varepsilon,\varepsilon)= \prod_{k=1}^{K} v_{k}^{\varepsilon-1}\mathrm{exp}(-\varepsilon v_{k}),
\end{equation}
where $\varepsilon>0$ is a small number which ensures the noninformativeness of the prior model,  $\mathbf{i}_{v}=[-\varepsilon\mathbf{1}_K,(\varepsilon-1)\mathbf{1}_K]$ and $\mathbf{i}_{\mu}=[-\varepsilon\mathbf{1}_K,(\varepsilon-1)\mathbf{1}_K]$ are natural parameters, { which are used to describe the distributions of varibles $v$ and $\mu$.}

Then, let us consider the squared error term $\|\mathcal{Y}-\left[\!\left[\mathbf{A}_1,\mathbf{A}_2,\cdots\mathbf{A}_d,\mathbf{X}\right]\!\right]\|_F^2$ in problem (\ref{problem2}). As  $\mathcal{N}$ is additive white Gaussian noise, we can interpret $\mathcal{Y}$ as the following negative log of a likelihood function
\begin{equation}\label{py}
	p\big(\mathcal{Y}|\mathbf{X},\beta\big) \propto \mathrm{exp} \Big(-\beta \|\mathcal{Y}-\left[\!\left[\mathbf{A}_1,\mathbf{A}_2,\cdots\mathbf{A}_d,\mathbf{X}\right]\!\right]\|_F^2\Big),
\end{equation}
where $\beta$ is the noise precision that can be modeled as gamma distribution, i.e.,
\begin{equation}\label{pvbeta}
	p\big(\beta|\mathbf{i}_{\beta}\big) = \prod_{k=1}^{K} \mathrm{Gamma}(\beta|\varepsilon,\varepsilon)= \prod_{k=1}^{K} \beta^{\varepsilon-1}\mathrm{exp} (-\varepsilon \beta),
\end{equation}
with natural parameter $\mathbf{i}_{\beta}=[-\varepsilon,\varepsilon-1]$.

Finally, We use $\mathbf{\Theta}$ to denote the aggregation of all the unknown random variables, that is, $\mathbf{\Theta}=\{\mathbf{X},\beta,\{\mu_k\}_{k=1}^K,\{v_k\}_{k=1}^K\}$. Combining (\ref{pG})-(\ref{pvbeta}), we can obtain the following joint pdf of $\mathbf{\Theta} $ and $\mathcal{Y}$ by using Bayesian formulas
\begin{equation}\label{joint}
	\begin{aligned}
	&p\big(\mathbf{\Theta},\mathcal{Y}\big) = p\big(\mathcal{Y}|\mathbf{X},\beta)\cdot p\big(\mathbf{X}|\{{\mu}_k\}_{k=1}^K,\{{v}_k\}_{k=1}^K)\cdot \\
	&p\big(\{{\mu}_k\}_{k=1}^K|\mathbf{i}_\mu)\cdot p\big(\{{v}_k\}_{k=1}^K|\mathbf{i}_v)\cdot p\big(\beta|\mathbf{i}_\beta) \\
	&\propto \mathrm{exp}\Big\{-\beta\|\mathcal{Y}-  \left[\!\left[\mathbf{A}_1,\mathbf{A}_2,\cdots\mathbf{A}_d,\mathbf{X}\right]\!\right]\|_F^2+(\prod_{i=1}^{d}l_i M)\ln\beta\\ &-\varepsilon\beta+(\varepsilon-1)\ln\beta -\mathrm{Tr}[\mathbf{\Lambda}(\mathbf{X}-\mathbf{P})^H(\mathbf{X}-\mathbf{P})]-\sum_{k=1}^{K}{\varepsilon v_k}\\
	&+(M+\varepsilon-1)\sum_{k=1}^{K}{\ln v_k}+(\varepsilon-1)\sum_{k=1}^{K}{\ln \mu_k}-\sum_{k=1}^{K}{\varepsilon \mu_k}\Big\}
	\end{aligned},
\end{equation}
where $\mathbf{\Lambda}=\mathrm{diag}(v_1,v_2,\cdots,v_K)\in\mathbb{C}^{K\times K}$ and $\mathbf{P}=[\mu_1^{-1}\mathbf{1}_M,\mu_2^{-1}\mathbf{1}_M,\cdots \mu_K^{-1}\mathbf{1}_M]\in\mathbb{C}^{M\times K}$. With the joint pdf $p\big(\mathbf{\Theta},\mathcal{Y})$ in (\ref{joint}), we apply the Bayesian inference method to learn the unknown variables in $\mathbf{\Theta}$ from the tensor data $\mathcal{Y}$ by maximizing the posterior distribution of $\mathbf{\Theta}$, i.e., $p(\mathbf{\Theta}|\mathcal{Y})$, which can be computed as
\begin{equation}\label{opdis}
	p(\mathbf{\Theta}|\mathcal{Y})=\frac{p(\mathbf{\Theta},\mathcal{Y})}{p(\mathcal{Y})}=\frac{p(\mathbf{\Theta},\mathcal{Y})}{\int p(\mathbf{\Theta},\mathcal{Y})d\mathbf{\Theta}}.
\end{equation}

\subsection{Algorithm Design}
The joint pdf $ p\big(\mathbf{\Theta},\mathcal{Y}\big)$ we derive in (\ref{joint}) is excessively complex, which prohibits us to get an exact Bayesian inference algorithm based on the posterior distribution in (\ref{opdis}) for the unknown parameters \cite{tensor1}. Particularly, it is intractable to compute the posterior distribution with multiple integrations in (\ref{opdis}). In order to deal with this problem, we apply the variational inference method which constructs a variational distribution $q(\mathbf{\Theta})$ to approximate the true posterior distribution $ p(\mathbf{\Theta}|\mathcal{Y}) $. To achieve this goal, $q(\mathbf{\Theta})$ can be regarded as the solution which minimizes the Kullback-Leibler (KL) divergence, that is
\begin{equation}\label{KL1}
	\begin{aligned}
	&\mathop{\mathrm{minimize}}_{q(\mathbf{\Theta})}\quad \mathrm{KL}(q(\mathbf{\Theta})|p(\mathbf{\Theta}|\mathcal{Y})) \\
	&\triangleq\mathop{\mathrm{minimize}}_{q(\mathbf{\Theta})}\quad -\mathbb{E}_{q(\mathbf{\Theta})}\bigg\{\ln\frac{p(\mathbf{\Theta}|\mathcal{Y})}{q(\mathbf{\Theta})}\bigg\}.\\
	\end{aligned}
\end{equation}

Obviously, if there is no other constraints imposed on $q(\mathbf{\Theta})$, the KL divergence will come to 0 as $ q(\mathbf{\Theta})=p(\mathbf{\Theta}/\mathcal{Y})$, which leads us back to the original intractable posterior distribution in (\ref{opdis}). To handle the problem (\ref{KL1}), mean-field approximation \cite{byinfer} is employed as an useful method to get a tractable solution. For mean-field approximation, it requires an assumption that the variational pdf $q(\mathbf{\Theta})$ can be represented in a completely factorized form, as $q(\mathbf{\Theta})=\prod_{j=1}^{J}q(\mathbf{\Theta}_j)$, where $\mathbf{\Theta}_j$ is a set partition of $\mathbf{\Theta}$, that is, $\bigcup_{j=1}^J \mathbf{\Theta}_j=\mathbf{\Theta}$ and $\bigcap_{j=1}^J \mathbf{\Theta}_j=\varnothing$, and $J$ is the number of set partitions. With this factorization, problem (\ref{KL1}) can be rewritten as
\begin{equation}\label{KL2}
	\mathop{\mathrm{minimize}}_{\{q(\mathbf{\Theta}_j)\}_{j=1}^J}\quad -\mathbb{E}_{\{q(\mathbf{\Theta}_j)\}_{j=1}^J}\bigg\{\ln\bigg(\frac{p(\mathbf{\Theta}|\mathcal{Y})}{\prod_{j=1}^{J}q(\mathbf{\Theta}_j)}\bigg)\bigg\}.
\end{equation}
Noting that the structure of $ \{q(\mathbf{\Theta}_j)\}_{j=1}^J$ is a fully factorized form, which motivates us to use the block coordinate descent method to gain a suboptimal solution of (\ref{KL2}). Specifically, by fixing the rest variational pdfs $ \{q(\mathbf{\Theta}_j)\}_{j\neq i},\forall i$ except  $ q(\mathbf{\Theta}_j)$, $q(\mathbf{\Theta}_j)$ can be optimized as
\begin{equation}\label{KL3}
	\mathop{\mathrm{minimize}}_{q(\mathbf{\Theta}_j)}\int q(\mathbf{\Theta}_j)\Big( -\mathbb{E}_{\prod_{i\neq j}q(\mathbf{\Theta}_i)}\ln {p(\mathbf{\Theta},\mathcal{Y})}{+\ln q(\mathbf{\Theta}_j)}\Big)d\Theta_j.
\end{equation}

By solving the above optimization problem (\ref{KL3}), the optimal solution can be computed as \cite{byinfer2}
\begin{equation}\label{KL4}
q^*(\Theta_j)=\frac{\mathrm{exp}(\mathbb{E}_{\prod_{i\neq j}q(\mathbf{\Theta}_i)}\ln p(\mathbf{\Theta,\mathcal{Y}}))}{\int\mathbb{E}_{\prod_{i\neq j}q(\mathbf{\Theta}_i)}\ln p(\mathbf{\Theta},\mathcal{Y})d\Theta_j } , \forall j.
\end{equation}
Based on the optimal solution $q^*(\Theta_j)$ in (\ref{KL4}), we can derive a closed-form posterior update for variational pdfs of each unknown variable in $\mathbf{\Theta}$.

We concentrate on deriving the variational distribution of the $q(\mathbf{X})$, whose mean matrix is a tight approximation of the desired device state matrix $\mathbf{X}$. Yet, the likelihood function proposed in (\ref{py}) results in complex computation among the device state matrix $\mathbf{X}$, which makes the $q^*(\mathbf{X})$ difficult to derive. In order to overcome this difficulty, we define
$\mathcal{Y}(d+1)\in\mathbb{C}^{M\times l_1 l_2\cdots\l_d}$ as an unfolding operation for an $(d+1)$th-order tensor $\mathcal{Y}\in \mathbb{C}^{l_1\times l_2\times \cdots \times l_d\times M}$ along its $(d+1)$-th mode. By substituting (\ref{joint}) into (\ref{KL4}), with the property of matrix trace $\|\mathbf{A}\|^2_F=\mathrm{Tr}(\mathbf{A}\mathbf{A}^H)$ and only keeping the terms relevant to $\mathbf{X}$, we get
\begin{equation}\label{qX}
	\begin{aligned}
	&q^*(\mathbf{X})\propto\bigg\{\mathbb{E}\bigg[-\beta\Big\|\mathcal{Y}-\left[\!\left[\mathbf{A}_1,\mathbf{A}_2,\cdots\mathbf{A}_d,\mathbf{X}\right]\!\right]\Big\|_F^2\\
	&-\mathrm{Tr}\Big(\mathbf{\Lambda}(\mathbf{X}-\mathbf{P})^H(\mathbf{X}-\mathbf{P})\Big)\bigg]\bigg\}  \\
	&\propto\mathrm{exp}\{-\mathrm{Tr(\mathbf{X}\cdot\Big[\underbrace{\mathbb{E}[\beta]\mathbb{E}[(\mathop{\diamond}\limits_{j=1}^d\mathbf{A}_j)^T(\mathop{\diamond}\limits_{j=1}^d\mathbf{A}_j)^*]+\mathbb{E}[\mathbf{\Lambda}]}_{\mathbf{C}^{-1}_{X}}}\Big]\cdot\mathbf{X}^H \\
	&-\mathbf{X}\mathbf{C}^{-1}_{X}\bigg[\bigg(\underbrace{\mathbb{E}[\beta]\mathcal{Y}(d+1)\Big(\mathop{\diamond}\limits_{j=1}^d\mathbb{E}[\mathbf{A}_j]\Big)^*+\mathbb{E}[\mathbf{\Lambda}]\mathbb{E}[\mathbf{P}]\bigg)\mathbf{C}_X}_{\mathbf{M}_{X}} \bigg]^H\\
	&-\mathbf{M}_{X}\mathbf{C}_X^{-1}\mathbf{X}^H\},
\end{aligned}
\end{equation}
where $\mathop{\diamond}\limits^d_{j=1}\mathbf{A}_j= \mathbf{A}_1\diamond\mathbf{A}_2\diamond\cdots\diamond\mathbf{A}_d $ denotes the multiple Khatri-Rao products. It is found that the device state matrix $\mathbf{X}$ obeys the circularly symmetric complex matrix Gaussian distribution $\mathcal{CN}_{M\times K}(\mathbf{X}|\mathbf{M}_X,\mathbf{1}_{M}\otimes \mathbf{C}_X)$ with mean matrix $\mathbf{M}_X$ and covariance matrix $\mathbf{1}_{M}\otimes \mathbf{C}_X$.

As mentioned above, $\mathbf{M}_X$ can approximate the desired device state matrix $\mathbf{X}$. Thereby, we focus on the derivation of $\mathbf{M}_X$. As seen in (\ref{qX}), $\mathbf{C}_X$ is involved in $\mathbf{M}_X$. In this context, we derive all terms in $\mathbf{M}_X$ and $\mathbf{C}_X$ in the following.

Firstly, we calculate the expectation $\mathbb{E}[\mathbf{\Lambda}]$ consisting of $\mathbb{E}[{v}_k], k=1,2,\cdots,K$. By substituting (\ref{joint}) to (\ref{KL4}) and removing the terms irrelevant to $v_k$, we remain
\begin{equation}\label{qv}
	\begin{aligned}
		&q^*(\{v_k\}_{k=1}^K)\propto\mathrm{exp}\{\mathbb{E}[-\mathrm{Tr}(\mathbf{\Lambda}(\mathbf{X}-\mathbf{P})^H(\mathbf{X}-\mathbf{P}))\\
		&-\sum_{k=1}^{K}\varepsilon v_k+(M+\varepsilon-1)\sum_{k=1}^{K}\ln v_k]\},
	\end{aligned}
\end{equation}
which is equivalent to $q^*(\{v_k\}_{k=1}^K)=\prod\limits_{k=1}^{K}q^*(v_k)$ with
\begin{equation}\label{qv2}
	\begin{aligned}
	&q^*(v_k)\propto\mathrm{exp}\\
	&\{-v_k\underbrace{\mathbb{E}[\mathbf{X}(:,k)^H\mathbf{X}(:,k)-2\mu_k^{-1}(\sum_{m=1}^{M}\mathbf{X}(m,k))+M\mu_k^{-2}+\varepsilon]}_{a_{v_k}}\\
	&+(\underbrace{M+\varepsilon}_{b_{v}}-1)\ln v_k\}.
\end{aligned}
\end{equation}
From (\ref{qv2}), it is known that the optimal $q^*(v_k)$ obeys the gamma distribution as $\mathrm{Gamma}(v_k|a_{v_k},b_v)$. For the mean $a_{v_k}$, we have
\begin{equation}\label{avk}
	\begin{aligned}
	a_{v_k}=&\mathbf{M}_X(:,k)^H\mathbf{M}_X(:,k)+M\mathbf{C}_X(k,k)\\
	&-2\mathbb{E}[\mu_k^{-1}]\sum_{m=1}^{M}\mathbf{M}_X(m,k)+M\mathbb{E}[\mu_k^{-2}]+\varepsilon.
\end{aligned}
\end{equation}
Then, according to the property of gamma distribution, the expectation of parameter $v_k$ can be calculated by $\mathbb{E}[v_k]=b_v/a_{v_k}$ with $b_v=M+\varepsilon$.

In a similar way, let us think about the parameter $\mu_k$ to calculate expectations $\mathbb{E}[\mu_k^{-1}]$ and $\mathbb{E}[\mu_k^{-2}]$. After plugging the proposed joint pdf (\ref{KL4}) into (\ref{joint}) and only remaining the terms related to $\mu_k$, we get
\begin{equation}\label{mu1}
	\begin{aligned}
	q^*(\{\mu_k\}_{k=1}^{K})\propto&\mathrm{exp}\{\mathbb{E}[-\mathrm{Tr}(\mathbf{\Lambda}(\mathbf{P}^H\mathbf{P}-\mathbf{X}^H\mathbf{P}-\mathbf{P}^H\mathbf{X}))\\
	&+(\varepsilon-1)\sum_{k=1}^{K}{\ln \mu_k}-\sum_{k=1}^{K}{\varepsilon \mu_k}]\},
\end{aligned}
\end{equation}
Due to  $q^*(\{\mu_k\}_{k=1}^{K})=\prod\limits_{k=1}^{K}q^*(\mu_k)$, we obtain
\begin{equation}\label{mu2}
	\begin{aligned}
		q^*(\mu_k)\propto&\mathrm{exp}\{-\mu_k^{-2}\underbrace{M\mathbb{E}[v_k]}_{o_{\mu_k}}+\mu_k^{-1}\Big(\underbrace{2\mathbb{E}[\sum_{m=1}^{M}\mathbf{X}(m,k)]\mathbb{E}[v_k]}_{t_{\mu_k}}\Big)\\
		&-\varepsilon\mu_k+(\varepsilon-1)\ln \mu_k\}
	\end{aligned}	
\end{equation}
There exist two additional terms $\mu_k^{-2}$ and $\mu_k^{-1}$ in the pdf (\ref{mu2}) compared with gamma distribution, which make the calculation of $\mathbb{E}[\mu_k^{-1}]$ and $\mathbb{E}[\mu_k^{-2}]$ much more complicated. To this end, we expect to obtain an approximation of (\ref{mu2}). In order to approximate (\ref{mu2}), the term $\varepsilon\mu_k$ is neglected since it is sufficiently small. Therefore, the expectation of $\mu_k$ in (\ref{mu2}) can be derived. For simplicity, we use $a_{\mu_k}$ and $b_{\mu_k}$ to denote the coefficient of $\mu_k^{-2}$ and $\mu_k^{-1}$. In other words, we have $a_{\mu_k}=M\mathbb{E}[v_k]$ and $b_{\mu_k}=2\sum\limits_{m=1}^{M}\mathbf{M}_X(m,k)\cdot\mathbb{E}[v_k]-\varepsilon$. In this case, $\mathbb{E}[\mu_k^{-1}]$
and $\mathbb{E}[\mu_k^{-2}]$ can be cast as (\ref{emuk}) and (\ref{emuk2}) at the top of next page.
\begin{figure*}
\begin{align}\label{emuk}
\mathbb{E}[\mu_k^{-1}]=\frac{t_{\mu_k}\cdot\mathrm{Gamma}[1-\frac{\varepsilon}{2}]\mathrm{Hy}[1-\frac{\varepsilon}{2},\frac{3}{2},\frac{t_{\mu_k}^2}{4o_{\mu_k}}]+\sqrt{o_{\mu_k}}\cdot\mathrm{Gamma}[\frac{1-\varepsilon}{2}]\mathrm{Hy}[\frac{1-\varepsilon}{2},\frac{1}{2},\frac{t_{\mu_k}^2}{4o_{\mu_k}}]}{o_{\mu_k}\cdot\mathrm{Gamma}[-\frac{\varepsilon}{2}]\mathrm{Hy}[-\frac{\varepsilon}{2},\frac{1}{2},\frac{t_{\mu_k}^2}{4o_{\mu_k}}]+\sqrt{o_{\mu_k}}t_{\mu_k}\cdot\mathrm{Gamma}[\frac{1-\varepsilon}{2}]\mathrm{Hy}[\frac{1-\varepsilon}{2},\frac{3}{2},\frac{t_{\mu_k}^2}{4o_{\mu_k}}]},
\end{align}
\end{figure*}
\begin{figure*}\begin{align}\label{emuk2}
\mathbb{E}[\mu_k^{-2}]=\frac{\sqrt{o_{\mu_k}}\cdot\mathrm{Gamma}[1-\frac{\varepsilon}{2}]\mathrm{Hy}[1-\frac{\varepsilon}{2},\frac{1}{2},\frac{t_{\mu_k}^2}{4o_{\mu_k}}]+t_{\mu_k}\cdot\mathrm{Gamma}[\frac{3-\varepsilon}{2}]\mathrm{Hy}[\frac{1-\varepsilon}{2},\frac{3}{2},\frac{t_{\mu_k}^2}{4o_{\mu_k}}]}{o_{\mu_k}^{\frac{3}{2}}\cdot\mathrm{Gamma}[-\frac{\varepsilon}{2}]\mathrm{Hy}[-\frac{\varepsilon}{2},\frac{1}{2},\frac{t_{\mu_k}^2}{4o_{\mu_k}}]+o_{\mu_k}t_{\mu_k}\cdot\mathrm{Gamma}[\frac{1-\varepsilon}{2}]\mathrm{Hy}[\frac{1-\varepsilon}{2},\frac{3}{2},\frac{t_{\mu_k}^2}{4o_{\mu_k}}]}.
\end{align}
\end{figure*}

Finally, to derive the expectation $\mathbb{E}[\beta]$, the posterior distribution of noise precision $\beta$ is updated by the following equation
\begin{equation}\label{beta}
    \begin{aligned}
    		q^*(\beta)\propto&( \underbrace{(\prod_{i=1}^{d}l_iM)+\varepsilon}_{b_{\beta}}-1)\ln\beta\\
    		&-\beta\underbrace{\mathbb{E}\Big[\|\mathcal{Y}-  \left[\!\left[\mathbf{A}_1,\mathbf{A}_2,\cdots\mathbf{A}_d,\mathbf{X}\right]\!\right]\|_F^2+\varepsilon\Big]}_{a_\beta}.
    \end{aligned}
\end{equation}
Checking $q^*(\beta)$ in (32), it is easy to identify $q^*(\beta)=\mathrm{Gamma}(\beta|a_{\beta},b_{\beta})$. Note that in (\ref{beta}), $b_{\beta}$ is related to the number of dimensions and $a_{\beta}$ estimates the residual of model fitting measured by the squared Frobenius norm. In order to calculate $\mathbb{E}[\beta]$, namely $a_{\beta}$ in (\ref{beta}), we unfold the tensor and then expand the Frobenius norm as follows
\begin{equation}\label{abeta}
	\begin{aligned}
	\mathcal{F}=&\mathbb{E}\Big[\big\|\mathcal{Y}-  \left[\!\left[\mathbf{A}_1,\mathbf{A}_2,\cdots\mathbf{A}_d,\mathbf{X}\right]\!\right]\big\|_F^2\Big]\\
	&=\mathrm{Tr}\Big(\mathop{\odot}\limits_{i=1}^d\big(\mathbf{A}_i^H\mathbf{A}_i\big)^H\times\big(\mathbf{M}_X^H\mathbf{M}_X+M\mathbf{C}_X\big)^H  \\
	&-\mathbf{M}_X\Big(\mathop{\diamond}_{i=1}^d\mathbf{A}_i\Big)^T\mathcal{Y}(d+1)^H\\
	&-\mathcal{Y}(d+1)\Big(\mathop{\diamond}_{i=1}^d\mathbf{A}_i\Big)^*\mathbf{M}_X^H\Big)+\big\Vert\mathcal{Y}(d+1)\big\Vert_F^2,
	\end{aligned}
\end{equation}
where (\ref{abeta}) holds true due to the fact of
$(\mathop{\diamond}\limits^d_{j=1}\mathbf{B}_j)^T(\mathop{\diamond}\limits^d_{j=1}\mathbf{B}_j)^* =\mathop{\odot}\limits^d_{j=1}\mathbf{B}_j^T\mathbf{B}_j^*$ [30], where $\mathop{\odot}\limits^d_{j=1}\mathbf{B}_j^T\mathbf{B}_j^*= (\mathbf{B}_1^T\mathbf{B}_1^*)\odot(\mathbf{B}_2^T\mathbf{B}_2^*)\odot\cdots\odot(\mathbf{B}_d^T\mathbf{B}_d^*)$ denotes the multiple Hadamard products. Similarly, for the term $\mathbb{E}\Big[(\mathop{\diamond}\limits_{j=1}^d\mathbf{A}_j)^T(\mathop{\diamond}\limits_{j=1}^d\mathbf{A}_j)^*\Big]$ in $\mathbf{C}_X^{-1}$ of (\ref{qX}), it can be reduced to
\begin{equation}\label{pro1}
	\mathbb{E}\Big[(\mathop{\diamond}\limits_{j=1}^d\mathbf{A}_j)^T(\mathop{\diamond}\limits_{j=1}^d\mathbf{A}_j)^*\Big]=\mathop{\odot}\limits^d_{j=1}\mathbb{E}\Big[\mathbf{A}_j^T \mathbf{A}_j^*\Big],
\end{equation}
Herein, to calculate the expectation on the right side of equation (\ref{pro1}), we provide the following theorem.

\emph{Theorem 1}: If $\mathbf{S}$ obeys the matrix-variate Gaussian distribution $\mathbf{S}\sim\mathcal{CN}_{M\times K}(\mathbf{S}|\mathbf{M}_S, \mathbf{C})$ with mean matrix $\mathbf{M}_S$ and covariance matrix $\mathbf{C}$. Then we have
\begin{equation}\label{pro21}
	\mathbb{E}\big[\mathbf{S}^H \mathbf{S}\big]=\mathbf{M}_S^H \mathbf{M}_S+\sum_{i=1}^{M}\mathbf{C}_{i,i},
\end{equation}
where $\mathbf{C}_{i,j}$ is the $(i,j)$-th block of $\mathbf{C}$.
\begin{IEEEproof}
	Please refer to Appendix A.
\end{IEEEproof}

Thereby, we obtain the statistics of all variational pdfs. It can be seen that the statistics of these variational pdfs are involved each other. In this context, these variables should be updated alternatingly until convergence. In summary, the tensor-based Bayesian learning algorithm for massive GF-RA in LEO satellite IoT can be described as Algorithm 1.

\begin{breakablealgorithm}
	\caption{: Tensor-Based Bayesian Learning Algorithm for JADCE}
	\label{alg1}
	\begin{algorithmic}[1]
		\STATE{\textbf{Input:}  $\mathcal{Y}$, $\{\mathbf{A_i}\}_{i=1}^d $ and total number of iterations $T$ }
		\STATE{\textbf{Output:}  $\mathbf{M}_X$}
		\STATE{\textbf{Initialization} $\mathbf{M}_{X}^{(0)}$, $\mathbf{C}_{X}^{(0)}$, $\alpha_{\beta}^{(0)} $, $\{\alpha_{v_{k}}^{(0)}\}_{k=1}^K$, $\{a_{\mu_{k}}^{(0)}, b_{\mu_{k}}^{(0)}, \mathbb{E}_{\mu_{k}^{-1}}^{(0)}, \mathbb{E}_{\mu_{k}^{-2}}^{(0)}\}_{k=1}^K$, iteration index $t=0$;}
		\REPEAT
		\STATE{
			\textbf{update the parameters of} $q(\mathbf{X})^{(t)}$:
			\begin{equation}\label{CXt}
				\begin{aligned}
	        \mathbf{C}_X^{(t)} =& \Big(\frac{b_{\beta}}{a_{\beta}^{(t-1)}}\mathop{\odot}\limits_{i=1}^{d}\big((\mathbf{A}_i^{(t-1)})^H\mathbf{A}_i^{(t-1)}\big)^*\\
	        &+\mathrm{diag}\big(\frac{b_v}{a_{v_1}^{(t-1)}},\frac{b_v}{a_{v_2}^{(t-1)}},\cdots\frac{b_v}{a_{v_K}^{(t-1)}}\big)\Big)^{-1}
	    \end{aligned}
\end{equation}
	\begin{equation}\label{MXt}
			\begin{aligned}
	\mathbf{M}_X^{(t)}&=\Big[ \frac{b_{\beta}}{a_{\beta}^{(t-1)}}\mathcal{Y}(d+1)\big(\mathop{\diamond}\limits_{i=1}^{d}\mathbf{A}_i^{(t-1)}\big)^*\\
	&+\mathbf{1}_{M\times1}[E_{\mu_1},E_{\mu_2},\cdots,E_{\mu_K}]\\
	&\times\mathrm{diag}\big(\frac{b_v}{a_{v_1}^{(t-1)}},\frac{b_v}{a_{v_2}^{(t-1)}},\cdots\frac{b_v}{a_{v_K}^{(t-1)}}\big)\Big]\mathbf{C}_X^{(t)}.  \\
\end{aligned}
\end{equation}
}
		\STATE{\textbf{update the parameters of}  $q(\mu_{k})^{(t)}$:
			\begin{equation}\label{amut}
			o_{\mu_k}^{(t)}=M*\frac{b_v}{a_{v_k}^{(t-1)}}
			\end{equation}
		\begin{equation}\label{bmut}
			t_{\mu_k}^{(t)}=2(\sum_{m=1}^{M}\mathbf{M}_X^{(t-1)}(m,k)\cdot a_{\mu_k}^{(t)})-\varepsilon
		\end{equation}
		}\begin{figure*}\begin{align}\label{Emuk}
		\mathbb{E}_{\mu_k^{-1}}^{(t)}=\frac{t_{\mu_k}^{(t)}\cdot\mathrm{Gamma}[1-\frac{\varepsilon}{2}]\mathrm{Hy}[1-\frac{\varepsilon}{2},\frac{3}{2},\frac{(t_{\mu_k}^{(t)})^2}{4o_{\mu_k}^{(t)}}]+\sqrt{o_{\mu_k}^{(t)}}\cdot\mathrm{Gamma}[\frac{1-\varepsilon}{2}]\mathrm{Hy}[\frac{1-\varepsilon}{2},\frac{1}{2},\frac{(y_{\mu_k}^{(t)})^2}{4o_{\mu_k}^{(t)}}]}{o_{\mu_k}^{(t)}\cdot\mathrm{Gamma}[-\frac{\varepsilon}{2}]\mathrm{Hy}[-\frac{\varepsilon}{2},\frac{1}{2},\frac{(t_{\mu_k}^{(t)})^2}{4o_{\mu_k}^{(t)}}]+\sqrt{o_{\mu_k}^{(t)}}t_{\mu_k}^{(t)}\cdot\mathrm{Gamma}[\frac{1-\varepsilon}{2}]\mathrm{Hy}[\frac{1-\varepsilon}{2},\frac{3}{2},\frac{(t_{\mu_k}^{(t)})^2}{4o_{\mu_k}^{(t)}}]}
	\end{align}\end{figure*}
\begin{figure*}\begin{align}
		\label{Emuk2}
\mathbb{E}_{\mu_k^{-2}}^{(t)}=\frac{\sqrt{o_{\mu_k}^{(t)}}\cdot\mathrm{Gamma}[1-\frac{\varepsilon}{2}]\mathrm{Hy}[1-\frac{\varepsilon}{2},\frac{1}{2},\frac{(t_{\mu_k}^{(t)})^2}{4o_{\mu_k}^{(t)}}]+t_{\mu_k}^{(t)}\cdot\mathrm{Gamma}[\frac{3-\varepsilon}{2}]\mathrm{Hy}[\frac{3-\varepsilon}{2},\frac{3}{2},\frac{(t_{\mu_k}^{(t)})^2}{4o_{\mu_k}^{(t)}}]}{(o_{\mu_k}^{(t)})^{\frac{3}{2}}\cdot\mathrm{Gamma}[-\frac{\varepsilon}{2}]\mathrm{Hy}[-\frac{\varepsilon}{2},\frac{1}{2},\frac{(t_{\mu_k}^{(t)})^2}{4o_{\mu_k}^{(t)}}]+o_{\mu_k}^{(t)}t_{\mu_k}^{(t)}\cdot\mathrm{Gamma}[\frac{1-\varepsilon}{2}]\mathrm{Hy}[\frac{1-\varepsilon}{2},\frac{3}{2},\frac{(t_{\mu_k}^{(t)})^2}{4o_{\mu_k}^{(t)}}]}
\end{align}
\end{figure*}
$\mathbb{E}_{\mu_k^{-1}}^{(t)}$ and $\mathbb{E}_{\mu_k^{-2}}^{(t)}$ are updated according to (\ref{Emuk}) and (\ref{Emuk2}) at the top of next page
		\STATE{\textbf{update the parameters of} $q(v_{k})^{(t)}$:
		\begin{equation}\label{avt}
			\begin{aligned}
				&a_{v_k}^{(t)}=\mathbf{M}_X^{(t-1)}(:,k)^H\mathbf{M}_X^{(t-1)}(:,k)+M\mathbf{C}^{(t-1)}_X(k,k)\\
				&-2\mathbb{E}_{\mu_k^{-1}}(\sum_{m=1}^{M}\mathbf{M}_X(m,k))+M\cdot \mathbb{E}_{\mu_k^{-2}}^{(t-1)}+\varepsilon
			\end{aligned}			
	\end{equation}
}
		\STATE{\textbf{update the parameters of} $q(\beta)^{(t)}$ :}
			\begin{equation}\label{abetat}
			a_{\beta}^{(t)}=\mathcal{F}^{(t)}+\varepsilon
		\end{equation}
		\STATE{$t=t+1$;}
		\UNTIL{convergence}
	\end{algorithmic}
\end{breakablealgorithm}

Once we obtain the output $\mathbf{M}_X$ from Algorithm 1, we can perform the active device detection based on a threshold. As the activity probability $p_a$ is usually small, we set the detection threshold as $\theta=M(r\mathrm{max}(\vert\vert\mathbf{M}_X(m,n)\vert\vert))^2$ {\cite{C1}}, where $\mathrm{max}(\vert\vert\mathbf{M}_X(m,n)\vert\vert)$ is the biggest magnitude of element in the estimated device state matrix $\mathbf{M}_X$ and $r$ is the ratio of the maximum channel coefficient to the minimum channel coefficient. Thus, the activity detection result $\hat{\alpha}_{k}$ of device $k$ is given by
 \begin{equation}\label{detection}
 	\begin{cases}
 		\hat{\alpha}_{k}=1,     \text{   if $\big\vert\big\vert\mathbf{M}_X(:,k)\big\vert\big\vert_F^2\geq\theta $}    \\
 		\hat{\alpha}_{k}=0,      \text{   if $\big\vert\big\vert\mathbf{M}_X(:,k)\big\vert\big\vert_F^2<\theta $}
 	\end{cases} .
 \end{equation}
Once the device $k$ is detected to active, the corresponding CSI can be estimated as
        \begin{equation}\label{ches}
        \hat{\mathbf{h}}_k=\mathbf{M}_X(:,k)/\sqrt{\xi_k}.
        \end{equation}

\subsection{Algorithm Analysis}
To gain further insights from the above proposed algorithm, this subsection discusses its convergence property and computational complexity.
 \begin{enumerate}
\item \emph{Convergence Property:} For the functional minimization of the KL divergence in (\ref{KL1}), it is non-convex over the mean-field family $q(\mathbf{\Theta})=\prod\limits_{j=1}^Jq(\Theta_j)$. However, it is convex with respect to a single variational pdf $q(\Theta_j)$ if the others $\{q(\Theta_i)|i \neq j\}$ are fixed \cite{convp}. Hence, the proposed algorithm, which updates the optimal solution for each $\Theta_j$, is a coordinate-descent optimization strategy in the functional space of variational distributions with each update of single unknown variable $q(\Theta_j)$ solving a convex problem. Consequently, this guarantees monotonic decrease of the KL divergence derived in (\ref{KL1}), and also the algorithm is guaranteed to converge to a stationary point.

\item \emph{Computational Complexity:} For each iteration of the proposed algorithm, {the computational complexity is measured in terms of matrix multiplications. That is, if $\mathbf{A}\in\mathbb{C}^{m\times n}$ and $\mathbf{B}\in\mathbb{C}^{n\times m}$, then the computational complexity of matrix multiplication $\mathbf{A}\mathbf{B}$ is $O(m^2 n)$. Furthermore, the computational complexity of the proposed algorithm costs}  $O((\sum_{I=1}^dl_i+M)K^2+(d+1)\prod\limits_{i=1}^dl_iMK)$. Therefore, the overall complexity of the proposed algorithm is about $O(T((\sum_{I=1}^dl_i+M)K^2+(d+1)\prod\limits_{i=1}^dl_iMK))$, where $T$ is the number of iterations needed for convergence. It can be seen that the complexity of the proposed algorithm scales polynomially with the total number of potential devices $K$. In order to verify the computational efficiency, we have compared the proposed algorithm with other typical JADCE algorithms in Table \ref{ccc}. It is seen that the proposed algorithm is more computational efficient.
\end{enumerate}
\begin{table}[h]

	\centering
	\caption{COMPUTATIONAL COMPLEXITY COMPARISON}\label{ccc}
	\begin{tabular}{|c|c|}
		\hline
		SOMP & $O(LKM+K^3 M^3)$ \\\hline
		AMP  & $O(LKM)$  \\\hline
		Algorithm in \cite{G1} & $O(K^4)$ \\\hline
		Proposed algorithm & \makecell[c]{$O\bigg(\left(\sum_{I=1}^dl_i+M\right)K^2+(d+1)$\\$\times\prod\limits_{i=1}^dl_iMK\bigg)$} \\\hline
		
	\end{tabular}
\end{table}

\emph{Remarks}: By exploiting the tensor structure of the received signal, we propose a simple but effective JADCE algorithm via variational Bayesian learning. Such an algorithm can be adaptive to the complex and dynamic LEO satellite IoT environment, i.e., device activity, channel condition and noise statistics. Therefore, it is appealing to LEO satellite IoT.

\section{Simulation Results}
In this section, we provide extensive simulation results to testify the performances of the proposed algorithm in LEO satellite IoT. The simulation parameters are set in Table \ref{Spa} according to 3GPP TR 38.811 and TR 38.821. Generally, we use the error probability ($P_e$) and normalized mean square error (NMSE) to measure the accuracy of activity detection and channel estimation respectively, where NMSE is defined as $\frac{||\hat{\mathbf{X}}-\mathbf{X}||_F^2}{||\mathbf{X}||_F^2}$ with $\hat{\mathbf{X}}$ being the estimate of device state matrix $\mathbf{X}$.
\begin{table}[h]
	
	\centering
	\caption{Simulation Parameters}\label{Spa}
	\begin{tabular}{|c|c|}
		\hline
        Parameter & Value\\\hline\hline
        Satellite orbit & LEO \\\hline
        Carrier frequency $f$& 30GHz  \\\hline
        Altitude of orbit $d_0$ & 1000km \\\hline
        Carrier bandwidth $B$  & 25MHz  \\\hline
        Satellite antenna gain $b_k$  & 20dBi  \\\hline
        Transmit gain to noise temperature $G_k/T$ & 34dB/K \\\hline
        Boltzmann's constant $\kappa$ & 1.38 $\times 10^{-23} $ J/m \\ \hline
        Rain fading mean $\mu_r$& -2.6dB \\\hline
        Rain fading variance $\sigma_r^2$ &  1.63dB \\\hline
        3dB angle & $0.4^{\circ}$ \\\hline
        Rician factor $\lambda$ & 8 \\\hline
        LOS component $||\mathbf{h}_k^{LOS}||^2$&$\mathcal{U}[0.6,0.7]$  \\\hline
        NLOS variance $\mathbf{v}_k^{NLOS}$&$\mathcal{CN}[0.2,0.25]$  \\\hline

		Number of iterations $T$ & $35$\\ \hline
		Number of antennas $M$ & $4-8$\\ \hline
		Transmit SNR &$ 0-30$ dB  \\\hline
		Preamble length $L$& $50-400$ \\ \hline
		Number of potential devices $K$ & $200-800$\\\hline
		Activity probability $p_a$& $0.05-0.5$  \\\hline
	\end{tabular}
\end{table}

\begin{figure}[!h]
 \centering
\includegraphics [width=0.5\textwidth] {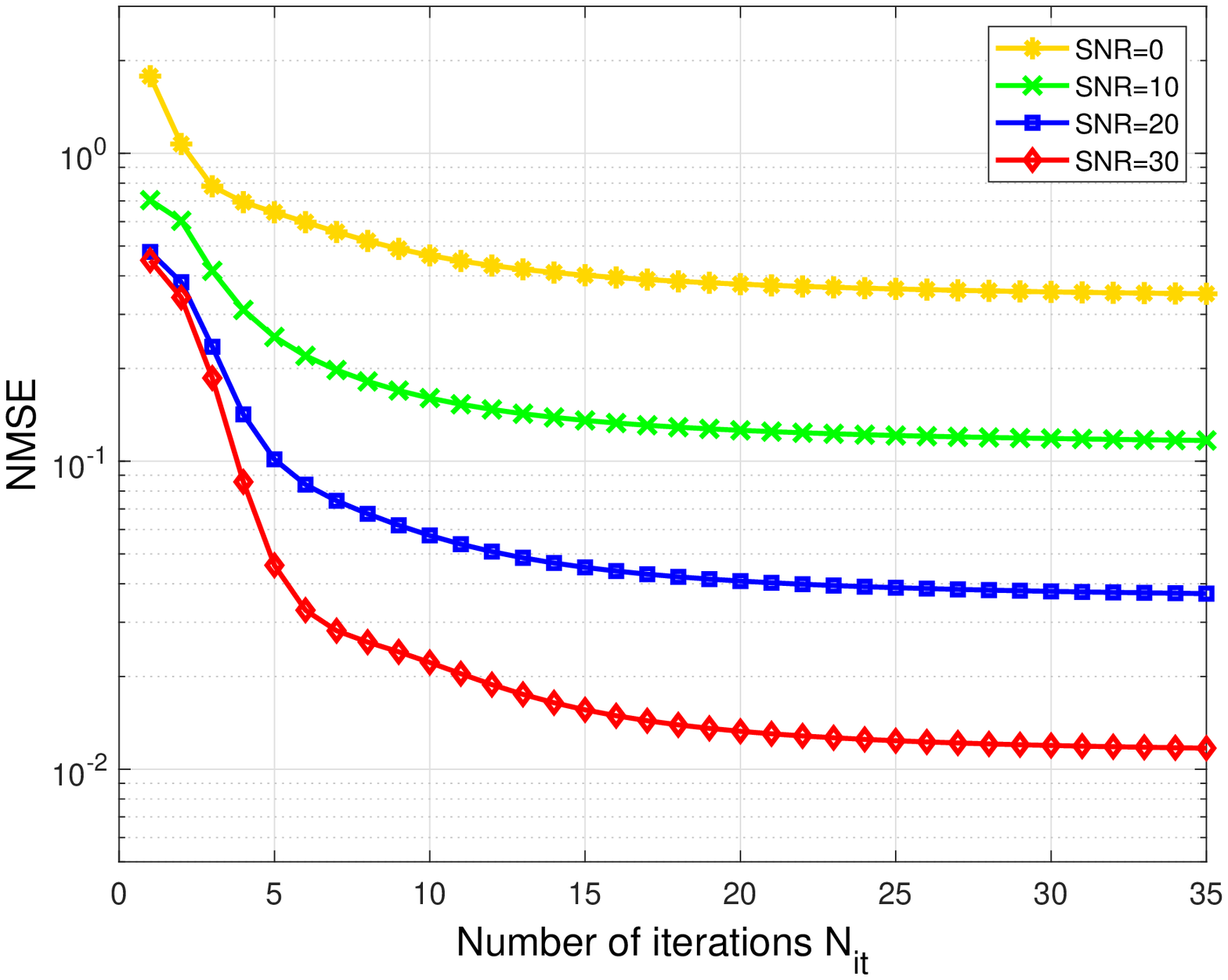}
\caption {NMSE performances under various SNR. Related parameters are $K=500, p_a=0.1, M=8, L=400.$}
\label{snr_figure}
\end{figure}

\begin{figure}[!h]
	\centering
	\includegraphics [width=0.5\textwidth] {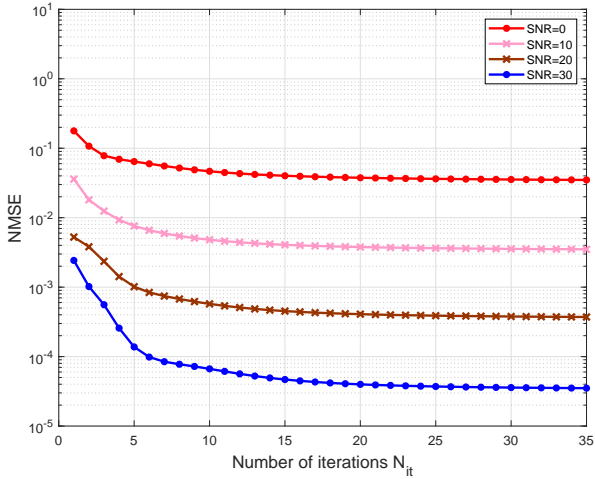}
	\caption {$P_e$ performances under various SNR. Related parameters are $K=500, p_a=0.1, M=8, L=400.$}
	\label{snr_pe}
\end{figure}

\subsection{Impacts of Signal-to-Noise Ratio}
Firstly, we validate the effectiveness of the proposed algorithm under various transmit signal-to-noise ratio (SNR), which is defined as $\mathrm{SNR}\triangleq10\log_{10}(\xi/\sigma_n^2$), where $\xi$ is the preamble transmit power and $\sigma_n^2$ is the noise variance. In general, IoT devices are required to use low transmit power, such that they can have a long life cycle.


It is shown in Fig. \ref{snr_figure} and Fig. \ref{snr_pe} that under various transmit SNRs, the proposed algorithm converges to a stationary point very fast. The required number of iterations is no more than 15. Hence, the proposed algorithm can be applied in LEO satellite IoT with time-varying environment. Moreover, with SNR=10 dB, it is possible to obtain low NMSE and $P_e$. As the transmit SNR increases, the NMSE and $P_e$ decrease accordingly. In other words, we can improve the accuracy of channel estimation by increasing the transmit SNR.

\subsection{Impacts of Tensor Decomposition Rank}
{
 In this subsection, we explore the impacts of tensor decomposition on the proposed JADCE algorithm. In general, the length of preamble $L$ can be factorized arbitrarily. However, for a given $L$, the number of factorizations $d$ affects the performance of the proposed algorithm due to different degrees of freedom (DoF) per active device. Specifically, since a variable in Grassmannian of lines in dimension $\tau_i$ has $\tau_i-1$ DoF \cite{D1}, the average of sum-DoF of the active devices in the model can be calculated as $\mathbb{E}\{\mathrm{DoF}(K,p_a)\}=Kp_a\sum_{i=1}^{d}(\tau_i-1)$. Therefore, given the total number of devices $K$ and acitivity probability $p_a$, the available DoF of $d=2$ is higher than that of $d=3$ and $d=4$, leading to better detection and estimation performance. This is also confirmed by simulation results. As seen in Fig. 5 and Fig. 6, with the increment of $d$, the NMSE and $P_e$ performance degrades. Yet, the increment of $d$ can reduce the computational complexity. Hence, we should choose a propoer $d$ according to the requirements of LEO satellite IoT.}

\begin{figure}[!h]
	\centering
	\includegraphics [width=0.5\textwidth] {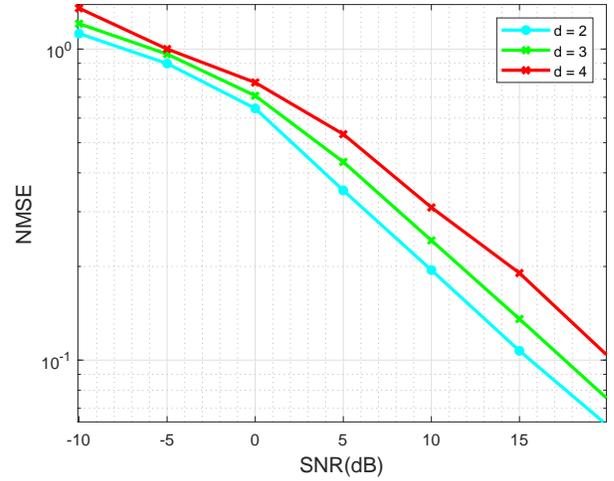}
	\caption {NMSE performances under various tensor decomposition orders $d$. Related parameters are $K=500, p_a=0.2, L=225, M=4.$ }
	\label{nmse_d}
\end{figure}

\begin{figure}[!h]
	\centering
	\includegraphics [width=0.5\textwidth] {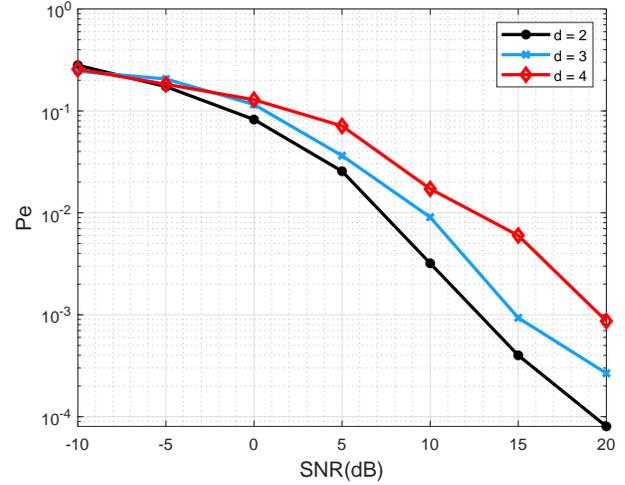}
	\caption {$P_e$ performances under various tensor decomposition orders $d$. Related parameters are $K=500, p_a=0.2, L=225, M=4.$}
	\label{pe_d}
\end{figure}

\subsection{Impacts of Preamble Length }

It is widely known that the preamble length $L$ has a great influence on the GF-RA performance of LEO satellite IoT. Given the requirements on NMSE and $P_e$, it is desired to use preamble sequences as short as possible, such that more duration in a time slot can be used for data transmission. Especially, for LEO satellite IoT, short packet is usually adopted to decrease the latency. In this context, it is necessary to improve the GF-RA performance with short preamble. In order to verify the superiority of the proposed algorithm, in this and following subsections, we will  compare it with three commonly-used JADCE algorithms, including Approximate Message Passing (AMP) algorithm \cite{AMP1}, Simultaneous Orthogonal Matching Pursuit (SOMP) algorithm \cite{OMP}, {and the algorithm in \cite{tensor1}.}

\begin{figure}[!h]
	\centering
	\includegraphics [width=0.5\textwidth] {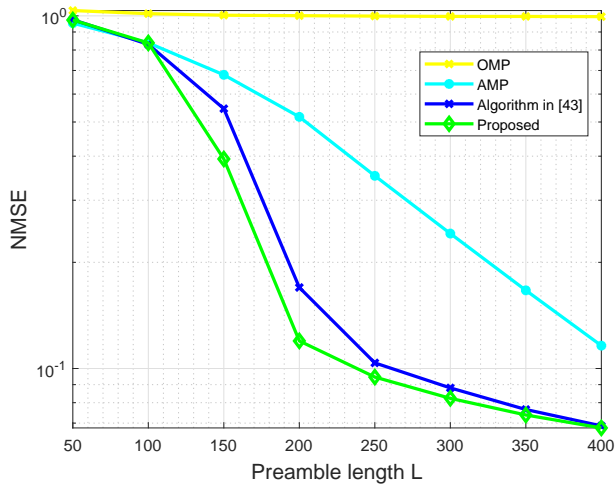}
	\caption {NMSE performances under various preamble length $L$. Related parameters are $K=500, p_a=0.2, \mathrm{SNR}=20\mathrm{dB}, M=4.$ }
	\label{nmse_L}
\end{figure}

\begin{figure}[!h]
	\centering
	\includegraphics [width=0.5\textwidth] {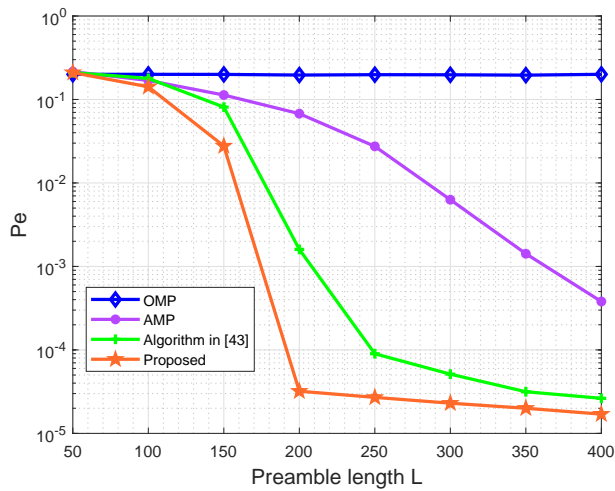}
	\caption {$P_e$ performances under various preamble length $L$. Related parameters are $K=500, p_a=0.2,\mathrm{SNR}=20\mathrm{dB}, M=4.$}
	\label{pe_L}
\end{figure}


As shown in Fig. \ref{nmse_L}, as the preamble length $L$ increases, the NMSE of the four JADCE algorithms decreases. In the whole preamble length region, the proposed algorithm performs best. Especially, as the preamble sequences become longer, the performance gain becomes larger. Similarly, for the $P_e$ performance shown in Fig. \ref{pe_L}, the proposed algorithm also performs best. For instance, at $P_e=10^{-2}$, the proposed algorithm can decrease the required preamble length about $120$ compared to the AMP algorithm. Thus, the proposed algorithm is suitable to LEO satellite IoT.

\subsection{Impacts of Activity Probability }
LEO satellite IoT needs to support various IoT applications in different scenarios, e.g., ocean, mountain, and desert. In general, these IoT applications may have quite different device activity probabilities. In this subsection, we compare the proposed algorithm and the other baseline algorithms with different activity probabilities.

\begin{figure}[!h]
	\centering
	\includegraphics [width=0.5\textwidth] {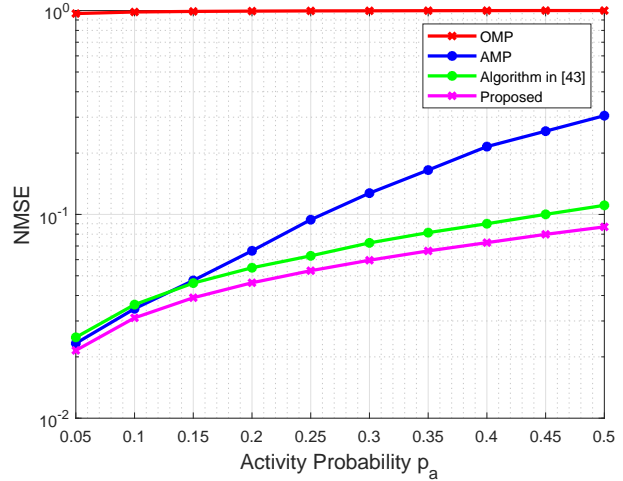}
	\caption {NMSE performances under various activity probabilities $p_a$. Related parameters are $K=500, L=400, \mathrm{SNR}=20\mathrm{dB}, M=8.$ }
	\label{NMSE_pa}
\end{figure}

\begin{figure}[!h]
	\centering
	\includegraphics [width=0.5\textwidth] {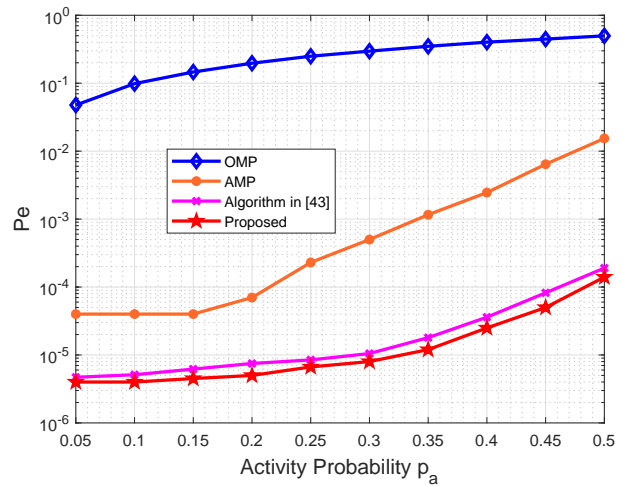}
	\caption {$P_e$ performances under various activity probabilities $p_a$. Related parameters are $K=500, L=400, \mathrm{SNR}=20\mathrm{dB}, M=8.$}
	\label{PE_pa}
\end{figure}

As seen in Fig. \ref{NMSE_pa}, for a given number of potential devices $K=500$, as the activity probability $P_a$ increases, the NMSE of the four JADCE algorithms increases. This is because the co-channel interference among active devices increases. The proposed algorithm still achieves the best performance, and the performance gain becomes larger as the activity probability increases. Moreover, it is shown in Fig. \ref{PE_pa} that the proposed algorithm has the lowest $P_e$, and obtains $P_e=10^{-4}$ even with $P_a=0.5$. Thus, the proposed algorithm can satisfy the requirements of various IoT applications.

\subsection{Impacts of Total Number of Potential Devices}
\begin{figure}[!h]
	\centering
	\includegraphics [width=0.5\textwidth] {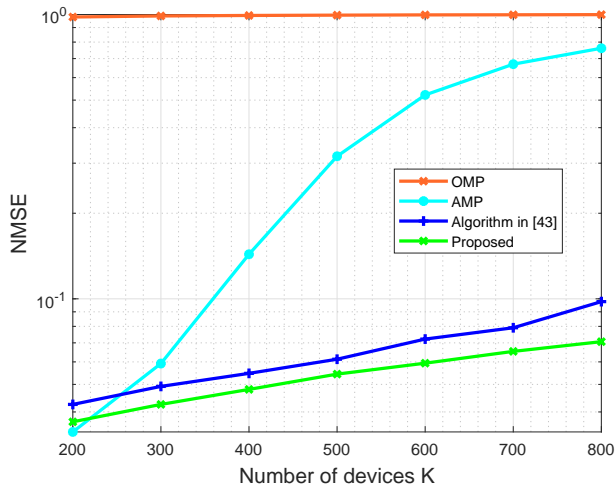}
	\caption {NMSE performances under various total numbers of potential devices $K$. Related parameters are $L=200, p_a=0.1, \mathrm{SNR}=20\mathrm{dB}, M=4.$} 
	\label{NMSE_K}
\end{figure}

\begin{figure}[!h]
	\centering
	\includegraphics [width=0.5\textwidth] {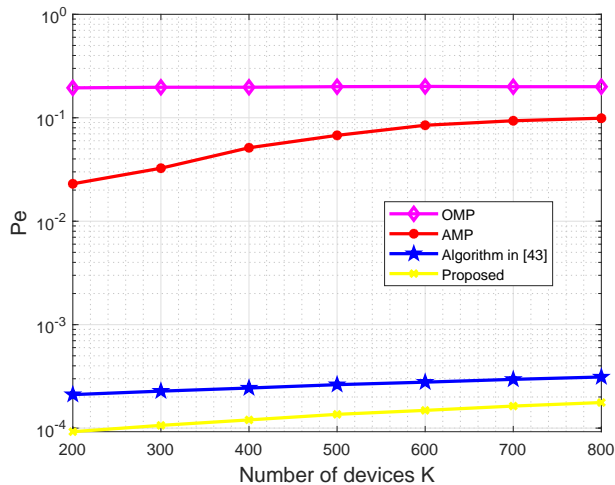}
	\caption {$P_e$ performances under various total numbers of potential devices $K$. Related parameters are $L=200, p_a=0.1, \mathrm{SNR}=20\mathrm{dB}, M=4.$ }
	\label{Pe_K}
\end{figure}

With the widespread applications of IoT, the number of IoT devices experiences an explosive increase. Hence, LEO satellite IoT must admit a massive number of potential devices. In this subsection, we examine the capability of the proposed algorithm in the sense of massive connectivity.
Fig. \ref{NMSE_K} shows the NMSE of the four JADCE algorithms with different numbers of potential devices for a given activity probability $p_a=0.1$. Intuitively, the NMSE of the four algorithms increases as the number of potential devices increases. Fortunately, the NMSE of the proposed algorithm increases very slightly when the number of potential devices increases from 200 to 800. Similarly, as shown in Fig. \ref{Pe_K}, the $P_e$ of the proposed algorithm is also not sensitive to the number of potential devices. Hence, the proposed algorithm is able to support massive connectivity.

In summary, the proposed algorithm can support low power, massive connectivity and wide coverage of IoT applications. Therefore, it is appealing to LEO satellite IoT.

\section{Conclusion}
In this paper, we have provided a novel massive GF-RA framework for LEO satellite IoT with low power, massive connectivity and wide coverage. By transforming the received signal to a tensor decomposition form, we proposed a Bayesian learning algorithm that can intelligently detect active devices and estimate channel state information. Both theoretical analysis and numerical simulations confirmed that the proposed algorithm had a low complex but good performance in LEO satellite IoT.

\begin{appendices}
\section{The Proof of Theorem 1}
According to (i) of Lemma 3.1 in \cite{appendix}, which proves that if random variable $\mathbf{y}$ follows the vector-valued Gaussian distribution $\mathcal{CN}_{MK}(\mathbf{y}|\bar{\mathbf{y}},\mathbf{\Sigma})$ with mean vector $\bar{\mathbf{y}}$ and covariance matrix $\mathbf{\Sigma}$, then we have $\mathbb{E}[\mathbf{y}(i)]=\bar{\mathbf{y}}(i)$ and $\mathbb{E}[\mathbf{y}(i)\mathbf{y}(j)]=\bar{\mathbf{y}}(i)\bar{\mathbf{y}}(j)+\mathbf{\Sigma}(i,j)$. In this way, we can get
\begin{equation}\label{expe}
    \mathbb{E}[\mathbf{X}(i,j)\mathbf{X}(m,n)]	=\mathbf{M}_X(i,j)\mathbf{M}_X(m,n)+\mathbf{\Omega}_{i,m}(j,n).
\end{equation}
Therefore, the result (\ref{pro21}) in \emph{Theorem 1} is proved.

\end{appendices}

\end{document}